# Blockchain and Cryptocurrency in Human Computer Interaction: A Systematic Literature Review and Research Agenda

Michael Fröhlich*
Center for Digital Technology and Management, Germany
froehlich@cdtm.de

Franz Waltenberger†
Center for Digital Technology and Management, Germany
waltenberger@cdtm.de

Ludwig Trotter
Lancaster University, w UK
l.k.trotter@lancaster.ac.uk

Florian Alt
Bundeswehr University Munich, Germany
florian.alt@unibw.de

Albrecht Schmidt
Ludwig Maximilian University, Germany
albrecht.schmidt@ifi.lmu.de

## ABSTRACT

We present a systematic literature review of cryptocurrency and blockchain research in Human-Computer Interaction (HCI) published between 2014 and 2021. We aim to provide an overview of the field, consolidate existing knowledge, and chart paths for future research. Our analysis of 99 articles identifies six major themes: (1) the role of trust, (2) understanding motivation, risk, and perception of cryptocurrencies, (3) cryptocurrency wallets, (4) engaging users with blockchain, (5) using blockchain for application-specific use cases, and (6) support tools for blockchain. We discuss the focus of the existing research body and juxtapose it to the changing landscape of emerging blockchain technologies to highlight future research avenues for HCI and interaction design. With this review, we identify key aspects where interaction design is critical for the adoption of blockchain systems. Doing so, we provide a starting point for new scholars and designers and help them position future contributions.

*Also with Ludwig Maximilian University, Bundeswehr University Munich.
†Also with Technical University of Munich.

## 1 INTRODUCTION

First introduced in 2008 as a peer-to-peer electronic cash system [97], blockchain technology has since drawn broad attention from research and industry alike. A growing body of literature envisions how its decentralized approach can disrupt current business models, financial systems, organizations, and civic governance [33, 34, 68, 121]. Arguably, the most visible evidence of growth is the combined market capitalization of over USD 1.7 trillion cryptocurrencies have accumulated by January 2022 [23]. Furthermore, developer activity has been steadily growing over the last decade [29], multiple projects have been started to improve over the original design (e.g. [15, 69, 138, 140]), and blockchain technology has been explored for a wide range of different applications and domains [35]. Through technical innovations, blockchains have advanced towards performance soon comparable to existing distributed systems – e.g. the Solana blockchain aims for a throughput of up to 710,000 transactions per second [140].

Despite these improvements, more than a decade after the launch of the Bitcoin network, blockchain technology seems to be far away from its envisioned omnipresence. In spite of avid calls from Human-Computer Interaction (HCI) scholars to engage with blockchain [35, 45], immature interaction concepts appear to hold back users with less technological affinity and present a barrier for wider adoption: Blockchain applications are hard to get started with [49, 52], confront both beginners and experienced users with misconceptions [87, 133], and are largely difficult to use [132]. While there have been systematic reviews of blockchain research in adjacent fields – e.g. security and privacy [144], current theories and models [58], and decentralized finance (DeFi) [92] – there is not yet a complete overview of HCI research pertaining to blockchain. To date, Elsden et al. arguably provide the most complete overview, yet without following a systematic approach and including only literature up to 2018 [35]. In a field characterized by rapid innovation, we thus see the need for a systematic review to understand the past, present, and future of HCI research on blockchain technology.

The objective of this paper is to develop an overview that can serve as a starting point when researching and designing with blockchain technology by showing how the field developed, mapping addressed questions and open challenges. To this end we ask the following research questions:

- How has HCI research on blockchain and cryptocurrency developed since the inception of Bitcoin?
- What themes, challenges, and design knowledge are discussed in the current research body?
- What are gaps that offer promising avenues for blockchain research in human computer interaction?

To address these questions, we conducted a systematic review of articles at the intersection of HCI and blockchain technology. We identified 99 relevant articles published between 2014 and 2021. While the majority has been published at SIGCHI conferences, there is a long tail of research published elsewhere. We organize the existing research body into six overarching themes and contrast them to the evolving blockchain ecosystem. Doing so, we highlight research opportunities for HCI and argue that interaction design research should boldly adopt modern blockchains as design materials to explore the creation of interactive decentralized applications.

**Contribution Statement:** With this systematic review, we make the following contributions: First, we present a descriptive overview of current blockchain and cryptocurrency research through an analysis of publication year, publishing databases, contribution types, and methodologies. Second, we analyze the existing research body and consolidate the produced knowledge into six major themes. Third, we conclude the paper by discussing salient gaps within the existing body of literature and draw up future research avenues for HCI and interaction design.

## 2 METHOD

The focus of this review is to summarize HCI related literature concerning cryptocurrency and blockchain. We structured the literature review in four overarching steps, following the PRISMA systematic review protocol [93]. An overview of our search results is depicted in Figure 1.

### 2.1 Step 1: Keyword Search

We selected the ACM Digital Library[1], IEEE Xplore[2], and Springer Link[3] as initial databases for this review. As a first step, we conducted a keyword search across all databases. We defined two sets of search terms: one related to the technology – i.e. blockchain – and one related to our research field – i.e. interaction design. The keywords were chosen by reviewing salient literature published at HCI venues (e.g. CHI, DIS, ToCHI) and iteratively refining them. Technology keywords[4] included for example "bitcoin", "cryptocurrency", and "blockchain". Qualifier keywords[5] included for example "user interface", "usability", and "design implication". We then computed query strings with pairwise combinations of technology and qualifier keywords and ran them against each of the databases. A publication would be included in our keyword search if either of the fields "title", "abstract", or "author keywords" matched against the pairwise combination. An abstract example of a query string would looks as follows: "*(title: keyword1 OR abstract: keyword1 OR author_keywords: keyword1) AND (title: keyword2 OR abstract:keyword2 OR author_keywords:keyword2)*". We conducted the search in July 2021 and did not restrict the search to a specific timeframe. We included all papers published before July 31, 2021 — the date of our search. The keyword search resulted in a total of 1,362 papers. Additionally, we iteratively conducted a forward search with Google Scholar[6] for all publications included in the review resulting in an additional 51 papers.

---

[1] https://dl.acm.org/ (last-accessed 2022-02-18)
[2] https://ieeexplore.ieee.org/ (last-accessed 2022-02-18)
[3] https://link.springer.com/ (last-accessed 2022-02-18)
[4] **Technology Keywords:** "bitcoin", "cryptocurrency", "crypto currency", "blockchain", "block chain", "distributed ledger", "dlt", "dapp", "crypto assets". (At the time of our search "web3" and "nft" did not return any relevant academic results and were therefore excluded. Given the recent rise of both concepts, future literature reviews may consider adding them.)
[5] **Qualifier Keywords:** "ui", "user interface", "interaction design", "ixd", "interaction", "user study", "usability", "ux", "user experience", "prototype", "interface" "interview study", "user-centered", "user-focused", "focus group", "HCI", "behavior", "end-user", "design implication", "design recommendation"

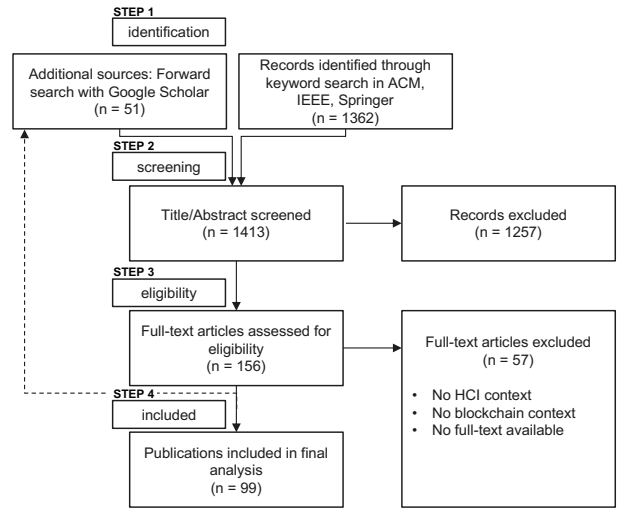

**Figure 1: PRISMA flow diagram of the screening process.**

### 2.2 Step 2: Screening Relevant Publications

In the second step, we screened the title and abstract of all 1,362 publications to identify those relevant for the review. We eliminated papers based on the following exclusion criteria:

- Publications with no blockchain or cryptocurrency focus
- Publications with no HCI focus (e.g. technical prototypes)
- Publications written in a language other than English
- Duplicates

A particularly high fraction of excluded publications can be attributed to keyword matches against "prototype" or "blockchain", resulting in technical prototypes without consideration of user interaction. In some situations, it was not apparent whether the exclusion criteria were met solely by looking at the title and abstract. In these situations, we included the publication for a full-text review in the next step to not miss relevant literature. In total 156 publications were reviewed – initially 105 to which 51 were added throughout the forward search process. All selected publications were downloaded for analysis in the next step.

---

[6] https://scholar.google.com/ (last-accessed 2022-02-18)

Table 1: Overview of the retrieved publications by year.

| Year | Total | Library (Sum) | | | | Publication Type (Sum) | | Metrics (Mean) | | |
| --- | --- | --- | --- | --- | --- | --- | --- | --- | --- | --- |
| | | ACM | IEEE | Springer | Other | Conference | Journal | Pages | Authors | Citations |
| 2014 | 1 | 0 | 1 | 0 | 0 | 1 | 0 | 5.0 | 4.0 | 37.0 |
| 2015 | 2 | 2 | 0 | 0 | 0 | 2 | 0 | 3.0 | 2.5 | 47.5 |
| 2016 | 3 | 2 | 1 | 0 | 0 | 3 | 0 | 7.7 | 3.7 | 33.7 |
| 2017 | 7 | 5 | 1 | 1 | 0 | 7 | 0 | 7.7 | 3.3 | 40.9 |
| 2018 | 14 | 11 | 2 | 0 | 1 | 12 | 2 | 10.2 | 4.0 | 25.9 |
| 2019 | 28 | 14 | 5 | 4 | 5 | 18 | 10 | 10.4 | 3.4 | 10.1 |
| 2020 | 26 | 10 | 6 | 8 | 2 | 22 | 4 | 9.6 | 4.5 | 4.1 |
| 2021 | 18 | 8 | 3 | 4 | 3 | 12 | 5 | 12.1 | 4.6 | 1.6 |
| | 99 | 52 | 19 | 17 | 11 | 77 | 21 | 8.2 | 3.7 | 25.1 |

*Notes.* Publications for the year 2021 are only included until July 31, 2021. Aggregated values are sums for the *Library* and *Publication Type* columns, and means for the *Metrics* columns. Citations numbers were retrieved from Google Scholar on December 20, 2021.

## 2.3 Step 3: Identifying Eligible Publications

In a final step, we reviewed the full text of the remaining publications. The eligible papers underwent more rigorous scrutiny based on the same exclusion criteria mentioned above, resulting in a final set of 99 papers.

## 2.4 Step 4: Qualitative Analysis

The 99 publications included in the review were read in full. In several iterations, the papers were analyzed and assigned codes. This information was entered into a database for further analysis. Throughout the process, publications were primarily coded by the main author and discussed for validation among the co-authors. Following a thematic analysis approach [12] the coded data was organized along initial emerging themes. In multiple rounds, these themes were revised, and the papers were re-coded until saturation was reached.

## 3 OVERVIEW

We included 99 publications in our review. Table 5 – located in the appendix – provides an overview of all included publications. For better accessibility, a spreadsheet of the table is included in the supplementary material. Table 1 provides an overview aggregated by *publication year*, *library*, *publication type*, and descriptive *metrics* of the papers. This review covers in total 8 years: The first included publication dates back to 2014, 6 years after the original Bitcoin whitepaper [97] was published. From then the number of publications increased year over year, peaking at 28 in 2019, slightly decreasing to 26 in 2020. These increases in scientific publications seem to be aligned with the crypto-hype-cycle peaks in 2013 and 2017, drawing in not only capital, startup activity, and developer activity [29], but as it appears also research interest.

The ACM Digital Library is the most relevant source with 52 (53%) publications, followed by IEEE and Springer. In total, eleven publications were identified from other databases (e.g. USENIX, Elsevier) using forward search. Only three venues have more than five publications attributed to them: CHI (21), DIS (7), and PACMHCI (5). A long-tail of 42 venues shows only one publication, indicating a fragmented field. Most work is published at conferences (78%), with journal publications only emerging over the past four years. The maturing of the field is also reflected by the steady increase of the average paper length (from 5.0 pages in 2014 to 12.1 pages in 2021) and the average number of authors contributing to a paper (from 4.0 authors in 2014 to 4.6 pages in 2021). The average paper has been cited 25.1 times. Not surprisingly, earlier publications show higher numbers of citations.

## 3.1 Two Perspectives: Blockchain or Cryptocurrency

We noticed that publications in our sample adopted one of two perspectives. Either they framed their research investigating *blockchain* technology (59, 60%) or *cryptocurrency* (40, 40%). Cryptocurrency publications mainly revolve around understanding users' motivation, perceived risks, and overall perception as well as users' interaction with wallets. Articles about blockchain focus on the design and development of blockchain-based systems for specific use-cases and their subsequent effects on users and society. The majority of empirical studies dealing with people evolve around cryptocurrency, whereas contributions about blockchain frequently contribute artifacts or system evaluations.

Among the 40 publications discussing cryptocurrencies in our corpus, 32 addressed Bitcoin, in eight cases the currency was not specified. This was, for example, the case when researchers explored the usability of different currency exchanges (e.g. [49, 64]). Among the 58 publications discussing blockchain, 13 addressed Bitcoin [97], 19 Ethereum [15], and six other blockchains such as IOTA [106]. 27 did not state a specific cryptocurrency. This was, for example, the case for publications surrounding interface prototypes (e.g. [11]) or design workshops (e.g. [32]).

## 3.2 Contribution Types

We coded all publications with regards to the contributions they were making, using the classification proposed by Wobbrock and Kientz [137] (see Figure 2). The majority of contributions are of empirical nature. In total 73 (74%) publications contribute either an *empirical study that tells us about how people use a system* (44 publications) or an *empirical study that tells us about people* (29 publications). 39 publications (38%) contribute an *artifact or system*. We included functional systems (e.g. [124, 129]) and interface or interaction prototypes (e.g. [9, 48]) under this category and excluded physical design kits (e.g. [72, 111]). Only few publications make *methodological* (2, 2%), *theoretical* (4, 4%), *dataset* (1, 1%), or systematic literature review (3, 3%) contributions. Eight publications (8%) contribute an *essay or argument*.

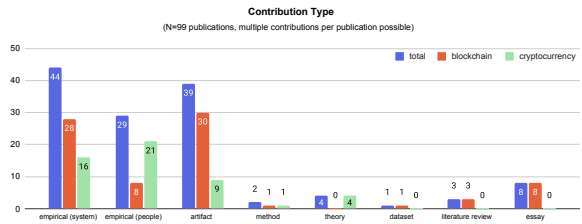

Figure 2: Contributions types made by publications in our sample.

## 3.3 Used Methods

We analyzed the research methods used across the included papers (see Figure 3). Several publications combined methods in their studies. We grouped the used data collection methods into six categories:

- **Quantitative data analysis** includes the analysis of secondary data such as log data (e.g. [41]), content analysis of forums and websites (e.g. [77]), or app reviews (e.g. [133]).
- **Interviews** include interview studies as primary source of data collection (e.g. [50, 51, 70, 115]) as well as interviews complementing evaluations of systems (e.g. [38, 79, 123]).
- **Questionnaires** include data collection through questionnaires as primary source of data collection (e.g. [2, 79]) as well as complementing other forms (e.g. [11, 143]).
- **Lab studies** include studies conducted in a lab environment in which rich data (e.g. screen-, video-, audio-recordings) could be collected. For example, usability studies (e.g. [8, 48, 104]) or heuristic evaluations through experts (e.g. [65]).
- **Field studies**, in contrast, include studies that are conducted in the natural environment of users. For example, ethnographic studies (e.g. [61, 62]) and deployed mobile applications (e.g. [11]) or systems (e.g. [40, 122]).
- **Workshops** include design research methods engaging groups of people in an effort to elicit design knowledge about people, specific systems, or speculative imaginaries (e.g. [37, 68, 111]).

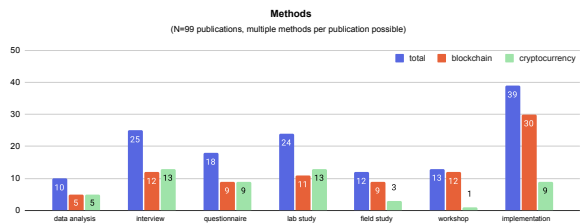

Figure 3: Method types used by publications in our sample.

The most used methods for data collection are interviews (25, 25%), lab studies (24, 24%), and questionnaires (18, 18%.) 51 publications report use of a single data collection method whereas 24 publications made use of method triangulation [105] by combining two or more types. For example, Tallyn et al. combined the analysis of log data and interviews during a field study deployment of an autonomous coffee machine [122]; Bidwell et al. used questionnaires and log data in a longitudinal field study to evaluate automated conditional giving [11]; Jabbar et al. used interviews and ethnographic observation to understand blockchain assemblages [62]. Looking at generative methods, there are several efforts to elicit design knowledge about blockchain systems in workshops, many of which make use of novel design kits [72, 88, 90, 111]. Most publications contributing artifacts – either in the form of interface prototypes or functional systems – present systems using blockchain to implement application-specific use cases (22 publications, e.g. conditional giving [129], energy trading [116], or last mile delivery [123]) or support tools (nine publications, e.g. visual smart contract construction [125], or tools for transaction analysis [75]).

## 4 MAJOR THEMES

After providing an overview of blockchain research in the HCI community, we present and discuss salient themes that emerged as we reviewed the papers. We identified 6 major themes: (1) the role of trust, (2) understanding motivation, risk, and overall perception of cryptocurrencies, (3) explorations surrounding the usability of cryptocurrency wallets, (4) engaging users with blockchain, (5) using blockchain for the implementation of specific use-cases, and (6) designing support tools for blockchain systems. Figure 4 visualizes the included publications over time per theme.

### 4.1 Trust in a Trustless System

A central feature of blockchain systems are their *trustlessness* – i.e. the fact that decentralized actors can agree on a common valid state of the systems without the need to trust a central entity or each individual actor within the system. Several HCI publications address trust and the trustworthiness of blockchain and cryptocurrency systems. This strand of research particularly challenges the assumption that blockchains are *trustless* and argues to adopt a sociotechnical perspective [25, 26, 53, 76, 82, 84, 116] as trust in algorithms cannot entirely substitute trust in humans [85]. Investigating the role of trust and how to design trustworthy systems is viewed as particularly important to understand the adoption or non-adoption by users [26, 131]. Figure 5 provides a visual overview.

*4.1.1 Factors Influencing Trust in Blockchain Systems.* Sas and Khairuddin were the first to integrate trust and blockchain in the context of HCI [70, 114, 115]. Drawing from established models of trust, they discuss the roles of technological trust, social trust, and institutional trust and conclude that established models fail to adequately address decentralized systems. They propose a research framework for HCI to explore trust along three layers and highlight users, merchants, miners, exchanges, and governments as relevant stakeholder groups for Bitcoin [114]. In the context of Bitcoin they define technological trust *as people's trust in Bitcoin technology experienced before, during, and after engaging in online transactions*, social trust as *the trust that Bitcoin stakeholders develop between*

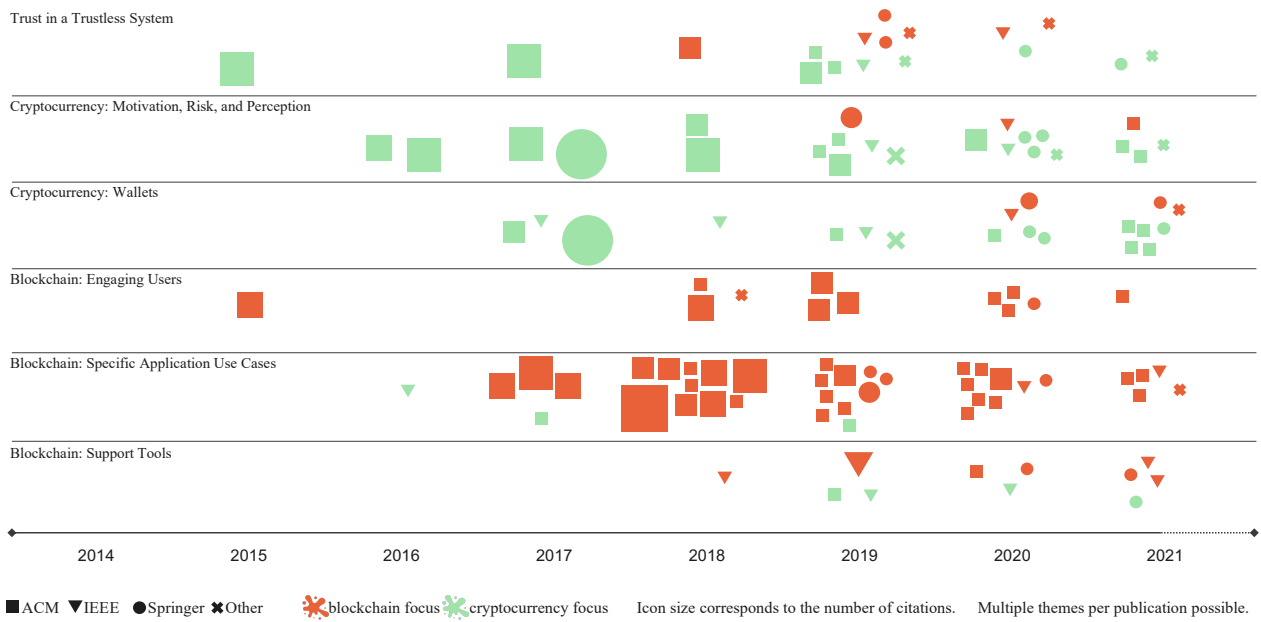

Figure 4: Overview of publications per major theme over time.

*each other*, and institutional trust *as the trust of governmental institutions in Bitcoin technology* ([114], p. 340). In subsequent work they explore both users' [115] and miners' trust perceptions [70] through qualitative interviews. The remaining papers subsumed under this theme primarily address end-users as stakeholder group. An exemption is the work by Voskobojnikov et al. who surveyed 204 non-users to investigate factors influencing the adoption of cryptocurrencies. Their results show that trust is a critical factor affecting adoption intention [131].

While Sas and Khairuddin's framework has found limited adoption among the sampled papers, we use it in the following to organize the trust building factors identified by research. Looking at factors that can be attributed to technological trust, we find several publications. Using a quantitative research design, Wallenbach et al. find that *immutability* and the *traceability of information* positively influence the trust in the technology. In contrast, the *anonymity of a blockchain* has a negative influence [134]. These results confirm the tension arising from having an open and decentralized, yet anonymous system, reported by Sas and Khairuddin [114]. Ooi et al. identify *technical protections*, *transaction procedures*, and *security statements* as determinants of perceived trust for Bitcoin [102]. Looking at social trust, Heidt identify *trust in code*, *in data*, in a *project's vision*, and *systemic trust* in the interplay between these factors to be relevant for design [53]. Craggs et al. emphasize the role of interpersonal trust in cryptocurrency communities, particularly *interpersonal trust in other users* and *interpersonal trust in the maintainer of the network* [26]. Additionally, several papers report the negative effect of illicit activities [115, 131] on trust in cryptocurrency systems. We did not identify any publications focusing on the trust relationship governmental institutions have towards cryptocurrencies or other blockchains. However, we noticed that a lack of trust in established institutions is a common theme mentioned by cryptocurrency users when asked why they are drawn to the space (e.g. [50, 71, 76, 79, 115]). Also dubbed "the paradox of unregulation", there are qualitative accounts arguing both for and against regulation of cryptocurrenies as a means to foster trust in cryptocurrencies [51, 70, 115, 132].

*4.1.2 Trust Challenges.* Grounded in the multifaceted factors identified to influence trust, scholars highlight different challenges. Between merchants and buyers, users face the *risk of dishonest traders* [115] as only one side of the transaction is recorded on the blockchain. The pseudonymous nature of transaction poses a challenge to establish trust over time. To mitigate this challenge social strategies are suggested (trade with authorized exchanges, socially authorized traders, reputable traders, or de-anonymized traders) and researchers call for technical advancements (e.g. to support two-way transactions and reversible transactions) [115].

We found that across several publications a lack of knowledge and experience of blockchain technology by most users is mentioned as reason for missing trust [20, 76, 82, 142]. Users with limited understanding have difficulties establishing (technological) trust [82]. For the adoption of centralized payment systems the reputation of the provider plays an important role (see e.g. [44] for Apple Pay). In the case of cryptocurrencies there is no central authority to trust. Because of that social elements gain importance, elevating, for example, the role of communities [116]. Knittel et al. report at the example of the Reddit r/bitcoin forum that the ideology within the community reduces interpretive complexity and supports collective imaginaries of a positive Bitcoin future [76].

On the technological side of the spectrum, *trust in data* remains an unsolved challenge. While data on the blockchain is immutable,

**Trust in a Trustless System**

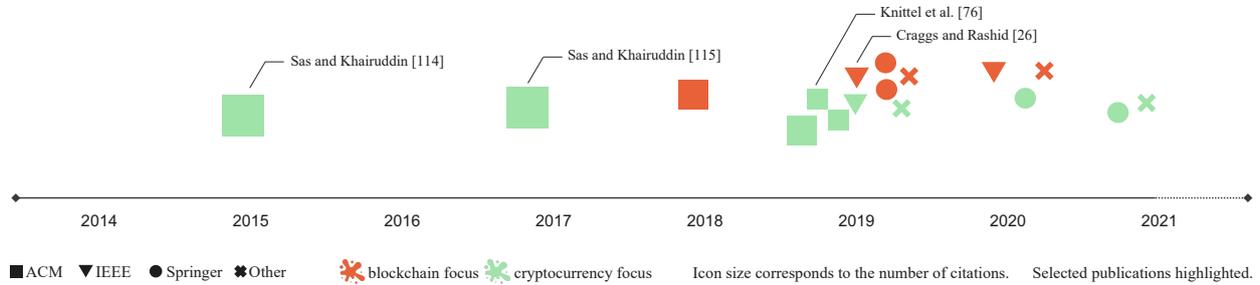

Figure 5: Overview of publications assigned to the *Trust in A Trustless System* theme.

the correctness of the data written on the blockchain cannot be verified easily (also known as oracle problem) [20]. Trust in reputable intermediaries to connect the real-world with the blockchain is thus necessary [20].

Finally, Bitcoin miners face additional trust challenges, specifically related to the fair distribution of mining rewards when contributing their mining power to a mining pool [70]. Beyond this we did not find other research addressing miners or validators.

*4.1.3 Designing Trustworthy Systems.* Several publications implement interfaces or functional systems to facilitate trust in blockchain systems. Lee et al. explore how a chatbot is used both as an object and mediator of trust and highlight the arising sociotechnical trust gap. At the example of the chatbot they argue that trust in a known technology (i.e. a chatbot interface) can mediate trust in an unknown technology (i.e. cryptocurrency) [82].

Drawing on the results of their quantitative study, Voskobojnikov et al. recommend to focus designing for *situational normality* to establish trust: Crypto-assets providers should mimic established payment systems users are already familiar with and provide stablecoins (cryptocurrencies that track the value of existing fiat currency) to lower the entry barrier [131]. Some scholars recommend the use of trust-supporting design elements in interfaces, such as trust-labels issued by known institutions such as exchanges [131], governments [142], or blockchain consortia [142].

### 4.2 Cryptocurrency: Motivation, Risk, and Perception

The second salient theme surrounds the exploration of the experiences and perceptions of cryptocurrency users. It is noticeable that publications in this cluster overwhelmingly focused on cryptocurrency users, not blockchain users. The large majority of publications focuses on Bitcoin and generalizes to cryptocurrencies. Figure 6 provides a visual overview.

*4.2.1 Motivation.* Several studies investigate the underlying motivation of why people are interested to engage with cryptocurrencies. While there is no established taxonomy, similar themes have been reported across studies. Froehlich et al. group users' motivation into *financial interest*, *ideological interest* and *technical interest* [50]. Abramova et al. present quantitative results separated by user groups, with *financial gain* and *interest in the technology* being the most important self-reported motives across all groups [2]. Similar motives are reported by Sas and Khairuddin [71, 115]: the *oncoming monetary revolution*, *empowerment associated with the use of a decentralized cryptocurrency*, *perceived material value*, and an *economic rationale*. Krombholz et al. report *curiosity* and the *decentralized nature* as motivators [79]. Voskobojnikov et al. take a different approach and investigate motivations and reasons against cryptocurrency adoption [131]. Contrary to qualitative reports by Gao et al. [51], they only find an indirect negative effect of self-efficacy on adoption intention. Among non-users, association with illicit activities, a lack of regulation, and the belief that Bitcoin's value has peaked were also reported to hold them back [132].

*4.2.2 Behavior and Perception.* Several studies attempt to increase knowledge on how cryptocurrency users are behaving and how their perception in turn influences behavior. There are multiple qualitative and quantitative studies reported. Common methods include questionnaires (e.g. [2, 79]), interview studies (e.g. [50, 51, 115]), and content analysis of forums and other data sources (e.g. [41, 76]). Quantitative studies provide insight into the demographic composition of cryptocurrency user base. Table 2 provides and overview. There are samples from different continents available. While the specific ratio shifts between studies, there are substantially more male participants than female ones. This skew is acknowledged by most authors, but we were not able to find any attempt explaining why women are less prevalent. The reliability of these demographic variables should be taken with a grain of salt as all studies adopt a targeted sampling procedure.

Table 2: Sample demographics of cryptocurrency users across quantitative empirical studies.

| Ref | Year | N | Geography | Age ($\mu$) | Gender (m/ f) |
| --- | --- | --- | --- | --- | --- |
| [2] | 2020 | 200 | US | – | 75% / 25% |
| [2] | 2020 | 195 | US, Canada, Europe | – | 80% / 15% |
| [102] | 2020 | 109 | Asia | – | 97% / 3% |
| [26] | 2020 | 125 | Europe, Americas | – | 88% / 12% |
| [79] | 2017 | 990 | US, Europe | 28.5 | 85% / 10% |
| [117] | 2014 | 134 | – | – | 95% / 5% |

*Notes.* All studies adopted a targeted sampling strategy.

**Cryptocurrency: Motivation, Risk, and Perception**

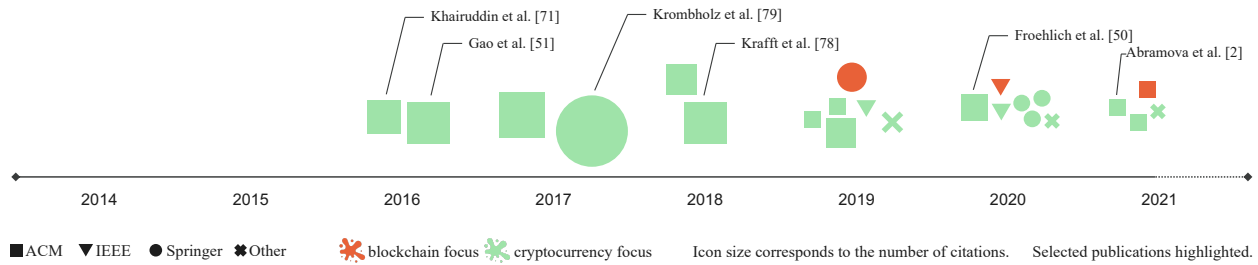

Figure 6: Overview of publications assigned to the *Cryptocurrency: Motivation, Risk, and Perception* theme.

Several studies report general usage behavior related to cryptocurrencies. Most do not distinguish between different cryptocurrencies or types of tokens; those that do limit their focus almost exclusively on Bitcoin. Users typically own more than one cryptocurrency [2, 79] and use different types of wallets in parallel [50] – a recent analysis of Abramova et al. shows that 80% have more than one type of wallet to manage their cryptocurrency [2]. Krombholz et al. provide additional insight into backup behavior [79]. With a mixed methods approach Busse et al. examine payment cultures in four countries (US, Germany, Iran, China), finding higher penetration of cryptocurrencies in western countries [14].

While Bitcoin is titled a *currency*, researchers have raised the question whether it is actually being used like one [89]. While Sas and Khairrudin report that most participants use Bitcoin primarily as store of value [115], Gao et al. find support for both investment and currency [51]. Froehlich et al. distinguish between use as *money* (i.e. as medium of transaction) and use as *asset* (i.e. as store of value or investment) and argue for designers to focus on either one use case to build more usable applications [50] .

Knittel et al. provide a deep qualitative analysis of the r/bitcoin community on Reddit[7] [76, 77]. They find that forum users subscribe to a "True Bitcoiner" ideology, consisting of three core beliefs: (1) *viewing Bitcoin's technology as more trustworthy than its people*, (2) *rejecting 'corrupt' social hierarchies related to money*, and (3) *the importance of accumulating or "HODLing" quantities of Bitcoin as a strategy to create an ideal future* ([76], p. 1). With a similar approach Jahani et al. try to disentangle processes of collective sense making related to emerging cryptocurrencies in forums [63]. Most Bitcoin users are not mining Bitcoin themselves [51, 79]. Khairuddin and Sas provide qualitative insights into the practices of Bitcoin miners, considering individual and collective approaches (solo-miners, collaborative mining pools, data-centers) [70].

Krafft et al. investigate how peer-influences affect user behavior on cryptocurrency exchanges. With a novel experimental approach they find that already low-value transactions affect buying behavior. They hypothesize about the role of user interface design elements (e.g. price history, tickers charts, price direction indicators) on collective behavior [78]. Being one of the few studies focusing on Ethereum, Faqir et al. analyze the effect of gas price surges on user activity in decentralized autonomous organizations (DAOs). Despite major surges in transaction fee costs in the analyzed time frame, they find only a minor influence on user activity [41].

*4.2.3 Risks, Security, and Privacy.* Connected to the overall perception of cryptocurrencies are the questions which risks users are exposed to, how they perceive them, and how they ultimately deal with them. These questions are particularly interesting in the context of blockchain systems, as many security-related tasks are shifted to the end user.

The most recent and arguably the most rigorous work surrounding risk perceptions and security behaviors of cryptocurrency users is presented by Abramova, Voskobojnikov, Beznosov, and Böhme [2, 131, 132]. Particularly, their CHI 2021 publication [2] is worth mentioning for three reasons. First, they connect to and synthesize 15 prior empirical studies offering an excellent starting point for new scholars in this field. Second, they thoroughly ground their study in theoretic underpinnings (the Protection Motivation Theory [112], the Theory of Planned Behavior [4], and the Technology Acceptance Model [27, 83]). And third, they combine a broad and deep sampling strategy to collect their data. Based on their survey results, they identify three distinct clusters of crypto-asset users – *cypherpunks, hodlers, and rookies*.

*Risks.* Engaging with cryptocurrencies requires users to deal with different risks. Abramova et al. surveyed cryptocurrency users about their perceived risk of being extorted, theft of private keys, loss through own mistakes, vulnerabilities of wallets, and vulnerabilities of exchanges [2]. Sas and Khairuddin highlight users' risks surrounding lost passwords, malicious attacks, dishonest trading partners, and failure to recover from human error or malice [115]. Building on their work, Froehlich et al. synthesize three essential risk categories: (1) *the risk of human error*, (2) *the risk of betrayal*, and (3) *the risk of malicious attacks* [50]. Across studies self-induced human errors are frequently reported (e.g. [2, 47, 50, 79, 87, 132]). Examples include forgotten passwords [115], forgotten storage locations, lost private keys, wrongly sent transactions [50], or ill investment decisions [2]. Risks of betrayal result from users misplacing trust in a third party [50], such as exchanges that fail to adequately protect their customers cryptocurrency. Examples for malicious behavior are also well documented: dishonest traders [115], extortion [2], theft [2, 79], and vulnerable wallets or exchanges [2].

---
[7]https://reddit.com/r/bitcoin (last-accessed 2022-02-18)

Interestingly, Voskobojnikov et al. find no significant effect of perceived risk on adoption intention. They reason that both users and non-users are most-likely aware of the most common risks [132]. Mai et al. find that while users are indeed able to explain a broad spectrum of risks, they often have incomplete or inaccurate mental models of how cryptocurrencies work [87]. Frequent misconceptions concern key management (who generates a key, how are transactions signed, that private keys should not be exposed) [87], what cryptocurrency addresses are [49, 87], transactions and fees (particularly how fees and transaction speed relate) [49, 50, 87, 133], and anonymity as well as security aspects [79, 87].

*Security and Privacy Personas.* Risk and security perceptions likely differ between individuals and it is reasonable to assume that cryptocurrency users are not a homogeneous group [2]. While studies try to distinguish between non-users [51, 131], beginners [48, 49], and cryptocurrency users [79, 102, 115], Abramova et al. are the first to define a typology of cryptocurrency users using an empirical approach [2]. They build on the concept of privacy personas [31, 80], a model distinguishing users based on their motivation and knowledge about security and privacy into five personas [31]. Froehlich et al. first connected privacy personas with user behavior in the cryptocurrency domain, suggesting that both knowledge and motivation about secure behavior would influence their risk perception. For example, *fundamentalists* (high knowledge, high motivation) would perceive a low risk of human error and value self-managed wallets over custodial ones. At the opposite side of the spectrum, the *marginally concerned* (low knowledge, low motivation) would prefer custodial wallets as they would perceive a higher risk of human error [50]. Abramova et al. applied hierarchical clustering on a sample of 395 participants and identified three robust clusters of users – *cypherpunks*, *hodlers*, and *rookies*. These personas differ in their security and privacy behavior. For example, cypherpunks rather opt for self-managed systems, whereas hodlers and rookies need to decide between custodial or self-managed wallets [2]. Their work may provide a valuable starting point for researchers who want to obtain a deeper understanding of user groups in cryptocurrency. Along with their analysis they also published the survey instrument they used to collect their data.

## 4.3 Cryptocurrency: Wallets

Wallets are the entrypoint to engage with blockchain applications and the cryptoeconomy at large.

We identified 16 publications which deal with the user experience or usability of wallets. Most publications present empirical results generated by evaluating one or several existing cryptocurrency wallets or exchanges [5, 8, 49, 64–67, 94, 109], or collected data through questionnaires [2, 79] and interview studies [50]. While most publications highlight challenges, usability issues, and provide recommendations to address them, hardly any implement and evaluate the proposed improvements. Surprisingly, only three publications contribute generative design artifacts: Froehlich et al. develop and evaluate onboarding flows to improve two wallets for beginners [48], Chen et al. present a prototype of an augmented reality cryptocurrency wallet [18], and Dlamini present a wallet for low cost mobile phones [30]. Beyond cryptocurrency wallets, we were surprised to find only one article focusing on decentralized applications (dApps) on the web [81]. Figure 7 provides a visual overview.

*4.3.1 Wallet Usability.* Several publications attempt to categorize wallets. Krombholz et al. initially present five categories related to key management and introduce the term "Coin Management Tool (CMT)" as synonym for wallet [79]. Froehlich et al. follow suit and distinguish between two categories: *Custodial wallets*, where a third party takes care of key management for users and *self-managed wallets* (also called non-custodial wallets [133]), where the user is in full control of and has full responsibility over key management [50]. Moniruzzaman et al. distinguish between mobile, hardware, paper, and web wallets [94]. In a similar fashion Voskobojnikov et al. distinguish software, mobile, hardware, paper, cloud, multi-signature, and brain wallets as well as exchanges [133]. Empirical studies reveal that most users have several types of wallets [2, 50, 79]. Custodial wallets are generally believed to be less secure, but more convenient to use for beginners [50, 133]. Scholars recommend the use of software wallets which are connected to the internet for use cases with frequent interactions, and more secure self-managed wallets for the long term storage of larger amounts [39, 50, 79]. Studies in our sample address custodial wallets [48, 49, 67, 109], self-managed wallets [5, 50, 133], decentralized exchange [64–66], or do not explicitly distinguish between them [2].

Wallets on desktop devices [5, 67, 109] and on mobile phones [18, 30, 48, 64–66] are looked at. Two studies address both desktop and mobile devices [49, 94]. One study looks into the usability and security of a hardware wallet [5]. There are several studies which focus explicitly on beginners or new users [8, 48, 49, 67]. Additionally, some studies engage with participants without any prior experience [64–66]. Surprisingly, we have not found any studies that evaluate the usability of wallets longitudinally. Table 3 provides and overview of typical tasks used to evaluate cryptocurrency wallets in lab studies.

**Table 3: Typical tasks during usability evaluations of wallets.**

| Task | Reference |
| --- | --- |
| Creating a new account (including verification) | [8, 49, 64, 65] |
| Creating a new wallet | [8, 49, 65–67, 94] |
| Depositing money and/or buying cryptocurrency | [49, 64–66, 94, 109] |
| Receiving or sending cryptocurrency | [8, 64–66, 109] |
| Purchasing goods with cryptocurrency | [8, 49] |
| Reviewing the portfolio value | [8, 48, 49, 65, 66, 66] |
| Backing up and restoring the wallet | [8, 94] |

While cryptocurrency wallets at large have not been attested great usability [5, 8, 49, 59, 64, 67, 94, 133], there are also a few examples suggesting that it is not impossible to develop usable cryptocurrency wallets: The best performing wallet in the heuristic evaluation of Moniruzzaman et al. has a task success rate of 97.3% [94]. Froehlich et al. report a SUS score [13] of 70 for one evaluated custodial wallet [49] and are able to improve the perceived usability of another wallet by designing an onboarding process [48].

**Cryptocurrency: Wallets**

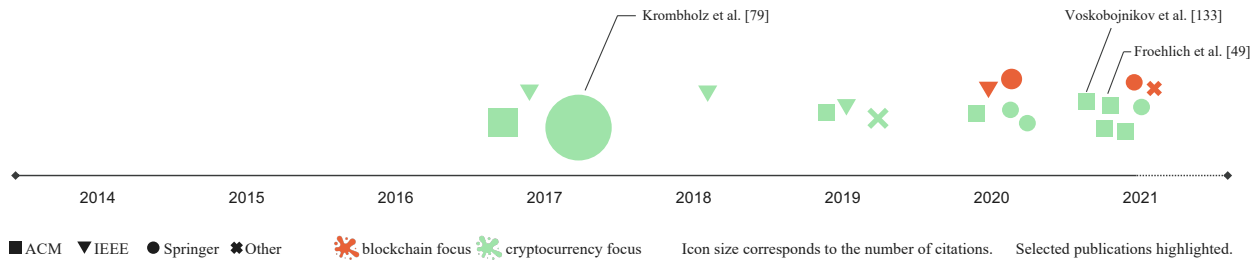

Figure 7: Overview of publications assigned to the *Cryptocurrency: Wallets* theme.

*4.3.2 Generalizable Design Insights.* Most publications present usability evaluations specific to the wallets they analyze [5, 18, 30, 64–67, 94, 109]. Only few publications aim at producing generalizable design insights about cryptocurrency wallets [2, 8, 48–50, 133]. For scholars new to the field, the most complete overview of usability challenges of cryptocurrency wallets is probably found in the works of Froehlich et al. [49] and Voskobojnikov et al. [133].

Froehlich et al. present the results of a qualitative user study with 34 novice users who engaged with custodial wallets for the first time. Using three different wallets, they identify several challenges that new users face when first interacting with cryptocurrencies and group them into three categories: *user interface challenges*, *finance challenges*, and *cryptocurrency challenges* [49]. The work by Voskobojnikov et al. complements these findings. They analyze app store reviews of self-managed wallets, identifying 6,859 reviews relating to user experience issues. Their thematic analysis suggests that both new and experienced users struggle with a range of issues: Confirming results from a similar analysis of finance apps [59], mobile cryptocurrency apps at large still suffer many shortcomings related to user experience [133]. Voskobojnikov et al. distinguish in their analysis between *General UX Issues* and *Domain Specific UX Issues* that are closer related to cryptocurrencies. We adopt this perspective to collate the design challenges and recommendations across the reviewed papers in the following.

*4.3.3 General User Experience Issues.* Across the analyzed papers there are many issues and shortcomings that are not unique to cryptocurrency wallets. While not unique, they become more severe given the direct involvement of money and the irreversible nature of cryptocurrency transactions [49, 133]. For example, Voskobojnikov et al. report a case where poor interface design resulted in direct monetary loss when a user sent a transaction multiple times [133]. Performance issues, freezes, crashes, outdated protocol implementations, and blocking user interfaces are being reported by app reviewers [132]. Different issues related to the structure and functionality of user interfaces are being reported across publications: Poor layout and structure of the interface [5, 49], ambiguous system status or inaccurate information [49, 133], and a general lack of guidance [49, 87, 133]. Additionally, issues pertaining to technical jargon [87, 94], confusing iconography and naming [49, 64, 133], typos [133], color schemes [133], and ill-designed error messages [49] are common. Another issue reported by Froehlich et al. in the context of custodial wallets concerns the extended sign-up or verification process, often required by regulation [49, 133]. They report that anti-money-laundering (AML) and know-your-customer (KYC) procedures often feel invasive for users, are error prone, disrupt the user experience through context and device switches, and sometimes lead to confusion about the legitimacy of an app [49].

The prevalence of these issues suggests that the below average usability of cryptocurrency apps (e.g. reported by [49, 59]) might only partly related to technical aspects of cryptocurrencies. This consequently means that many of these issues can be addressed by following established design guidelines [49, 133].

Voskobojnikov et al. emphasize the importance of error recovery [98, 133] and advise practitioners to design for *situational normality* by mimicking existing online banking or payment systems users are already familiar with [133]. Other scholars draw similar examples to existing finance applications [49, 64]. Additional recommendations include designing for transparency and control [87, 116], focusing on the promotion of cryptocurrenies' benefits [67], supporting users' learning experience [49, 116] and designing for fun use [51].

*4.3.4 Domain Specific User Experience Issues.* The second category of issues directly relates to the cryptocurrency domain. Issues under this category result either from the user interface and application design or from misconceptions of users. While the former can be addressed through better design, misconceptions can only be addressed by finding ways to educate users. Unfortunately, misconceptions about cryptocurrencies appear to be quite frequent [87, 133]. Studies with non-users and beginners have shown that cryptocurrencies are difficult to get started with [8, 49], also referred to as the *onboarding problem* [52]. Scholars attribute the steep learning curve, to the technology's embedded complexity [116] and complicated metaphors that often do not match users' expectations [49, 50, 132]. For example, several publications report confusion about the term "wallet" – drawing from their experience with physical wallets user expect different functionality [49, 50, 79] or they connect the term to other concepts such as the native iOS wallet app [49].

*Addresses.* Cryptocurrency addresses are another frequent cause for confusion among new users. Beginners regularly associate the term with e-mail addresses [49, 133]. Given that they are in essence long alphanumerical strings, it is not surprising that users find them difficult to handle [49] and hard to remember [64]. Almutari et al. show that this makes them vulnerable to man-in-the-middle attacks as they are difficult to compare [5].

*Cryptocurrency Valuation.* Several issues relate directly to cryptocurrencies' valuation. The high price volatility is reported to be problematic for everyday use [51, 115], particularly when making transaction and different platforms use different exchange rates [49, 133]. The often high exchange rates of cryptocurrencies (i.e. one Bitcoin being worth tens of thousands of US dollars) make them difficult to deal with. Users think in fiat currency when transacting [49], making it necessary to convert prices back and forth. When making purchases at everyday price points, the corresponding cryptocurrency value is a small sub-comma amount (i.e. 50 EUR would be 0.00089 BTC) that is hard to deal with [49]. Interestingly, all of these issues are being addressed on a technical level by so-called stable coins. To our knowledge, there is no published work that looks into the usability of stable coins.

*Transactions.* Being central to cryptocurrency wallets, many issues are reported relating to transactions. Interfaces that do not immediately show transactions after being sent, leave users in confusion about the state of the transaction [49, 64]. The status of pending transactions if often misunderstood [49, 133]. Resulting from an inaccurate mental model of how blockchains work [87], users often expect transactions to be reversible [115, 133]. With the majority of studies being conducted with Bitcoin, participants frequently report that they perceive transactions to be slow [49, 64, 115].

*Fees.* Fees emerged as another problematic and widely reported area: Many users have an incomplete or inaccurate understanding of fees [49, 87]. The relation between fees and transaction speed is unclear [87, 133], users often do not expect that they have to pay fees [49], and they are perceived as too high [133]. Wallet operators may charge additional platform fees making it even more complicated to understand fee structures [49, 64, 133]. Configuring transactions with too low fees can cause transactions to be stuck and not processed by miners and most interfaces do not offer the option to overwrite stuck transactions [133]. While some scholars recommend to simplify fee selection interfaces by providing expressive categories (i.e. "slow – low fees", "default", "fast – high fees" [87]), app reviews also show that some users take issue if they cannot configure fees themselves [133]. Fees calculated automatically based on heuristics were reported to be unexpectedly expensive if sent at unfortunate points of time [49].

*Ecosystem Integration.* Frequent tasks in the evaluation of cryptocurrency wallets involves the purchase of goods [8, 49]. While users would like to use them as a means of payment [50, 51], there is still a lack of mainstream adoption, making it difficult to find merchants [51]. Payment integrations that exist are perceived as problematic [49]. Froehlich et al. highlight the difficulties of using Bitcoin for online purchases when on a mobile device: While many wallets offer features to scan addresses displayed as QR code, this feature becomes useless when the QR code is displayed within the browser on the mobile device itself. Paired with missing shortcuts and broken links this makes it necessary to manually copy addresses and values back and forth [49]. They consequently argue for the necessity of better ecosystem integration to create a seamless checkout process [49], mimicking payment systems users are already familiar with [49, 133].

*Key Management.* Self-managed wallets largely expose the underlying technology and many users perceive dealing with key management as a burden and bad usability [50]. Some wallets generate key pairs without the knowledge of the user. While this can be perceived positively by users who do not want to deal with key management, it might be a restriction for others [94]. Given the often inaccurate understanding about key management [87], it might be negative in the long run to shield users of self-managed wallets from this. For example, many beginners do not know about the importance of their backup phrases [87] and users often struggle with recovery mechanisms of self-managed mobile wallets [133]. Given that irrecoverable keys are a frequent reason for cryptocurrency loss [79], scholars suggest different approaches. Mai et al. suggest to force users to input parts of their backup phrase to prove that they saved it [87]. Abramova et al. emphasize the importance of wallets to transparently communicate about key management, particularly about storage practices such as encryption [2].

*User Groups.* Across publications it is apparent that many wallets try to provide one-size-fits-all solutions [2, 50, 133]. However, both qualitative [49, 50, 133] and quantitative [2] studies provide evidence that cryptocurrency users are not a homogeneous group, but differ in their behavior and their needs. Scholars recommend to build wallets tailored to the needs of specific user groups [2, 49, 50, 133] and for different use cases [50]. Relevant dimensions for segmenting users have been identified in their security and privacy behavior and their affinity towards key management [2, 50]. To flatten the learning curve and enable beginners to get started, wallets should guide users through their cryptocurrency journey and create Aha! moments early on [48]. By allowing them to personalize their experience through user profiles [2], they can gradually progress from simple to more complex topics. The importance of educating users throughout this process is emphasized by many scholars [48, 49, 67, 116, 133], particularly to resolve misconceptions. This way, users might start with custodial wallets [50], learn about key management, and graduate to self-managed wallets [50, 133]

### 4.4 Blockchain: Engaging Users

Several papers in our review focus on engaging participants in workshops and design activities surrounding blockchain applications. These speculative formats make use of physical design kits or participatory design activities to either facilitate understanding about blockchain or elicit user-centered requirements for the development of systems. Figure 8 provides a visual overview.

*4.4.1 Engaging with Blockchain.* With blockchain being perceived as "black box technology" [88], we found several publications reporting workshops and methods to engage a broader audience in the exploration of the technology. Khairuddin et al. presented BlocKit, a teaching kit based on materials such as clay, paper and padlocks in order to demonstrate usage and materialize virtual concepts via physical objects [72]. Other researchers have used LEGO blocks and role-playing games featuring pizza-shaped learning materials to educate about blockchain-based systems [90, 111]. Reporting results from three workshops, Manohar and Briggs demonstrate how creative methods are useful to enable critical reflection and knowledge exchange about blackbox technologies. They argue that design workshops offer a useful bridge between disciplines and are a valuable resource to inform future oriented design implications [88]. Nissen et al. present GeoCoin, a functional location-based application for learning and speculative ideation with smart contracts, through which users explore urban debit and credit zones. Building on this experience, they invited participants to engage in the exploration and design of further use cases in a subsequent workshop format [100]. Finally, Kera et al. present a design fiction: They use "anticipatory prototyping" to explore autonomous governance and combine a technical prototype with the artistic design fiction of *Lithopia*, a village governed by smart contracts. In this fictional village drones execute smart contracts based on the visual detection of certain actions among villagers by drones and satellites. The ultimately goal of the project was to explore and challenge promise of automated smart blockchain governance of participants and "onlookers" [68].

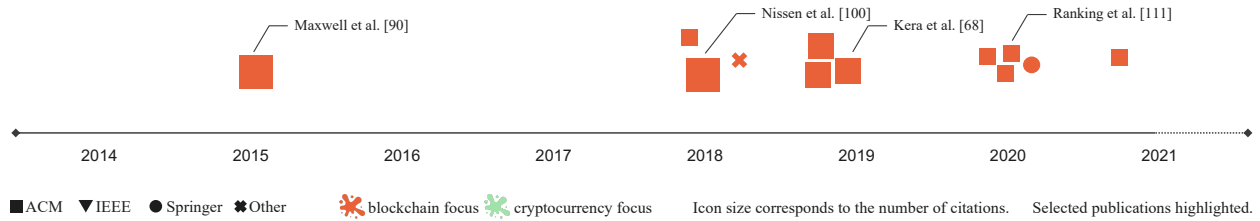

Figure 8: Overview of publications assigned to the *Blockchain: Engaging Users* theme.

*4.4.2 Participatory Design Activities.* We also identified multiple publications reporting participatory design activities with users. In contrast to the research summarized above, these papers aim at ideating specific use cases or eliciting design requirements from participants and less at helping participants better understand blockchain technology. Elsden et al. asked participants about their experiences with donating money and collected ideas and opinions on conditional donations [38]. Together with Oxfam they addressed a similar question from the perspective of charitable organizations, and explored potential use-cases with employees [37]. Others have, together with rural and urban agricultural communities, explored blockchain use cases to level environmental and social inequalities in food supply chains [55, 107]. Beyond these examples, participatory design approaches were used for exploring local energy trading systems [32], location-based blockchain applications [100], and smart-contract governed delivery scenarios [124].

## 4.5 Blockchain: Specific Application Use Cases

We identified 39 articles in our systematic review that propose or evaluate specific blockchain applications or use cases. Figure 9 provides a visual overview. We categorize these articles according to the topology of blockchain applications by Elsden et al. [35]. Articles with overlaps across the categories were assigned based on the article's main focus. An overview of our results can be found in table 4.

Table 4: Proposed systems in the application-specific use cases theme according to the typology by Elsden et al.

| Category | Count | Publications |
|---|---|---|
| Underlying Infrastructure | – | – |
| Currency | 4 | [40, 56, 60, 100] |
| Financial Services | 7 | [11, 20, 38, 107, 116, 128, 129] |
| Proof-as-a-service | 7 | [3, 37, 45, 61, 126, 136, 142] |
| Property and Ownership | 5 | [9, 19, 42, 54, 101] |
| Identity Management | – | – |
| Governance | 15 | [1, 16, 17, 22, 32, 34, 36, 55, 62, 84, 91, 113, 122–124] |

*Notes.* Articles with overlaps across the categories proposed by Elsden et al. [35] were assigned based on the article's main contribution. Elsden et al.'s paper [35] proposing the typology is not assigned as it discusses all categories equally.

*4.5.1 Underlying Infrastructure.* With blockchain protocols and decentralized ecosystems being the focus of more systems and cryptography oriented research, it is little surprise that this review found only a small number of articles focusing on underlying infrastructure technologies across research conducted in HCI. We identified work that uses blockchain technology as enabling, underlying software platform to create novel applications e.g. [1, 20, 38, 42, 129] and autonomous or semi-autonomous systems in the context of a networked internet of things [16, 17, 122]. While it may be argued that these examples fit into the taxonomy of underlying infrastructure, most of the work went beyond the mere technical implementation by exploring financial models, socio-economic phenomena and civic engagement and governance.

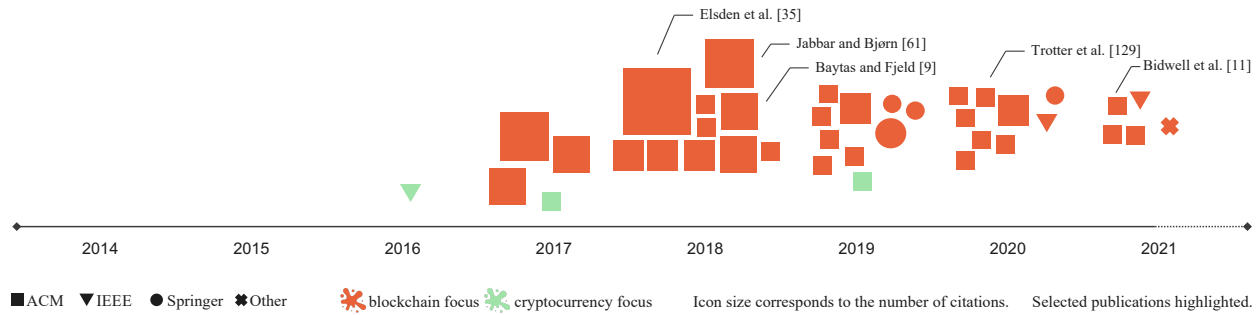

Figure 9: Overview of publications assigned to the *Blockchain: Specific Application Use Cases* theme.

*4.5.2 Currency.* Originally invented as a "peer-to-peer electronic cash systems" [97], digital currencies are still the most prevalent use case for blocking technology. In addition, cryptocurrencies and custom utility tokens not only find widespread use to facilitate the exchange of value in the majority of use-cases we revived (e.g. [1, 16, 17, 20, 38, 42, 42, 122, 129]), but form the underlying incentives for many to participate in the development and upkeep of the decentralized blockchain networks [97]. Sections 4.2 and 4.3 have covered work on motivations, risks and perceptions of digital cryptocurrencies and wallets, hence these are not taken into consideration in this section. Specific applications for currencies included an early point-of-sale (POS) system for a coffee shop to accept Bitcoin by Eskandari et al. [40], a browser plugin for tipping for educational resources [56], a prototype for mining cryptocurrency on mobile devices [60], and GeoCoin, an experimental platform enabling participants to interact with location-based smart contracts [100].

*4.5.3 Financial services.* A large body of HCI work focusing on financial services using on blockchain technologies developed around charitable donations. Research conducted by Elsden et al., Trotter et al., and Bidwell et al. [11, 128, 129] explored the use of blockchain technologies and smart contracts to increase trust and transparency through higher levels of agency and control. The "Smart Donations" system enables donors to attach rules to their charitable gift and triggers pre-specified pay-outs in response to real-world events that are validated through trusted third-party oracles [38, 128]. Trotter et al. outline domain considerations and challenges alongside a comprehensive reference implementation using smart contracts on the Ethereum blockchain. Notably, the authors decided to build a mobile application and custom user interface to abstract the underlying complexity of the Ethereum blockchain and highlighted challenges in the management and exchange of crypto-assets [129]. Their implementation was later evaluated by Bidwell et al. in an in-the-wild study with 93 donors over 8-weeks. The study provides insights into the temporal qualities that emerge from smart contracts that preserved and enforced financial intentions from donors. The authors suggest that sensitivity for time, when designing interactions with blockchains, could facilitate profound temporal orientations and meaningful user experiences [11]. Similarly, work by Chiang et al. demonstrate the potential of smart contracts as an automatic, impartial mediator to increase levels of trust among stakeholders in financial transactions. The authors find that for Mexican migrants living in the US, greater transparency and control around financial transactions and the flow of funds to their rural home communities facilitated by smart contracts can increase trust and cooperation between individuals and government institutions [20].

*4.5.4 Proof-as-a-service.* The use of blockchain technologies as a trusted digital data storage offers a plethora of possible use-cases and applications. While many applications make use of trusted digital storage on the ledger, often to facilitate higher degrees of trust [1, 20, 42, 107, 129], this section specific work developed around the theme of *proof-as-a-service*. Our review identified applications for provenance in supply and distribution chains, as a trustworthy, immutable digital notary for both, digital and physical artifacts and as immutable, trusted data registers.

We found many examples that investigated the application of blockchain technologies in supply and distribution chains. While some work has an emphasis on governance e.g. in agri-food [45, 107] and energy markets [32, 91, 116], Jabbar et al. provide detailed insight into the implementation of blockchain technology in the shipping industry [61]. Other work developed and evaluated a local courier service system based on smart contracts [123, 124]. Tharatipyakul and Pongnumkul [126] provide a comprehensive survey on user interfaces in blockchain-based agri-food provenance tracking applications. Their work categorizes means to collect (forms, scanning, and sensors) and visualize (text, tables, timelines, graphs, and maps) provenance data. Their work reveals usability challenges and emphasizes the need to consider interface design to widen blockchain adoption in the future [126].

Examples for blockchain in digital notaries include a reference architecture for an academic certificates registry [3] while [113] highlighted conflicts deploying such a system within a higher education institution. Using the example of a system that collects and stores the history of cars over their life cycle, Zavolokina et al. discuss trust-enhancing design elements that interaction and user interface designers can use to increase trust in blockchain-based proof-as-a-service applications [142]. Wenceslao and Estuar propose a hybrid system using hashed links between off-chain and on-chain storage to support secure, tamper-proof storage and access control of (audio) recordings of medical consultations [136].

*4.5.5 Property and Ownership.* With immutable and trustless digital ledgers, combined with enforceable rules governed by smart contracts, blockchains support applications that aim to proof, manage and enforce rights related to author- and ownership of all types of digital and physical assets [9, 35, 42]. Despite its significant potential, so far, only little research has been conducted in this space[8]. Baytas and Fjeld provide a design provocation challenging the notions of permanence and disposability of digital and physical artifacts, exploring how the traditional concept of passed-along-generations heirlooms can be transferred into the digital realm using blockchain technologies [9]. Chen and Ko suggest to use augmented reality do materialize digital pets owned on the blockchain [18]. OLeary et al. address the problem of social loafing in the workplace through a secure, transparent, immutable and verifiable system that captures ownership of an employees individuals intellectual property [101]. Fedosov et al. explore distributed ledgers in digital sharing economy services through a blockchain-enabled peer-to-peer lending system. Their "Just Share It" system enables individuals to share equipment (e.g. tools, sports gear, toys), aiming to disintermediate interactions, increase trust among peers and mediate claim management if borrowed items were damaged [42].

*4.5.6 Identity Management.* Self-sovereign identity management (SSI) is a well-known and widely researched use case that gained significant attention across academia [43, 96, 119], industry[9] and governments[10]. The European Union Agency for Cybersecurity recently released a comprehensive review of SSI [99] and pilot test of SSI technologies are currently being carried out in Germany[11]. Amid this cross-sector interest in self-sovereign identity management, our review has not yielded relevant research conducted in HCI to address interaction design challenges for identity management. The roleplay game, PizzaBlock, by Rankin et al. [111] touches on decentralized identity management for charity volunteers, albeit with a focus on educating non-technical users. Our findings highlight a research gap that should be actively addressed by the HCI and interaction design community in the future.

*4.5.7 Governance.* Elsden et al. highlight smart contracts' ability to facilitate distributed decision making and governance [35]. This section builds on their definition and provides an overview of HCI research that explores disintermediated control mechanisms, including semi-autonomous and autonomous systems and decentralized autonomous organizations (DAOs). Themes that emerged in our qualitative analysis of prior work included socio-technical challenges around autonomous human-machine interactions, new forms of organizational governance and community engagement.

Lustig discusses visions of decentralized autonomous systems and identifies three possible frames through which to interpret imaginaries about autonomous systems: (1) as physical objects, (2) as mathematical rules, or (3) as artificial mangers [84]. Tallyn et. al are the first to report the design of a blockchain-enabled system with the autonomously acting coffee machine BitBarista, which besides selling coffee was also capable of rewarding users for maintenance tasks such as replenishing beans or emptying coffee grinds [122] using Bitcoins. This idea was developed further by Cardenas and Kim which explored the design choices and social implications for financial robot-human agreements. Initial work presented roBU, a prototype robot that was able to provide financial incentives to humans helping the robot to archive targets (e.g. attending university classes and traveling around the world [16]. Later work included interactions with virtual robotic agents and more sophisticated configurations, e.g. an autonomous ride-sharing service [17].

Use-cases around organizational governance cover a broad scope. Several studies have discussed the use of decentralized smart contracts in the context of energy markets. Scuri et al. conducted human-centered research into self-governing, decentralized energy trading which provides insights into peoples perceptions, needs, motivations and proposes design guidelines for P2P energy trading platforms [116]. Doebelt and Kreußlein base their qualitative research on a similar use case exploring the needs and expectations of both consumers and considering gamification to facilitate engagement across the community. Notably, they conclude that energy supply through peer-to-peer communities should be considered as an additional rather than an alternative to the existing grid supply [32]. Early work by Meeuw et al. presents first results of user interface evaluations for autonomous peer-to-peer micro-grids [91].

Work by Rooksby and Dimitrov highlights the friction of deploying new forms of decentralized governance in established organizational structures by deploying a DAO within their university [113], while Abadi et al. aim to improve student engagement and participation through a decentralized student peer-trading platform with reputation system [1]. Other work explores the potential for socio-economic development and governance of rural communities through smart contracts. Pschetz et al. explore the use of decentralized governance in the context of smallholder farmers in the Caribbean. The authors highlight that the challenge is not in the actual money and commodity transactions but in the design of the terms and enforcement mechanisms implemented in smart contracts. [107]. This is developed further by Heitlinger et al., who discuss the possibilities of dehumanizing food systems through an algorithmic management on the blockchain.

## 4.6 Blockchain: Support Tools

We identified multiple publications which present support tools. While publications in the previous section used blockchain as design material to build systems, the ones presented here are auxiliary tools for blockchain [35]. The majority of publications in this category is not published in ACM, but in IEEE and Springer. Salient subtopics concern interactive tools to analyze and make sense of blockchain transaction data, as well as development support tools for smart contracts. Other prototypes include StockSense, a wrist-worn vibrotactile display that signals its users cryptocurrency market movements [104] and Brokerbot, a multiplatform cryptocurrency chatbot [82]. Figure 10 provides a visual overview.

---

[8] We are aware of recent research in the HCI community around the use of non-fungible tokens (NFTs) e.g. [46]. However, this research was conducted outside the time frame of this systematic review (see section 2) and has hence not been included in this review. We expect and encourage more work around the category of ownership and possession in the near future.
[9] https://www.typehuman.com/project/australian-red-cross (last-accessed 2022-02-18)
[10] https://idunion.org/ (last-accessed 2022-02-18)
[11] https://www.bundesregierung.de/resource/blob/998194/1898282/b5d50f1f53d99ee067edfcf43b2ecd31/digital-identity-neu-download-bundeskanzleramt-data.pdf (last-accessed 2022-02-18)

**Blockchain: Support Tools**

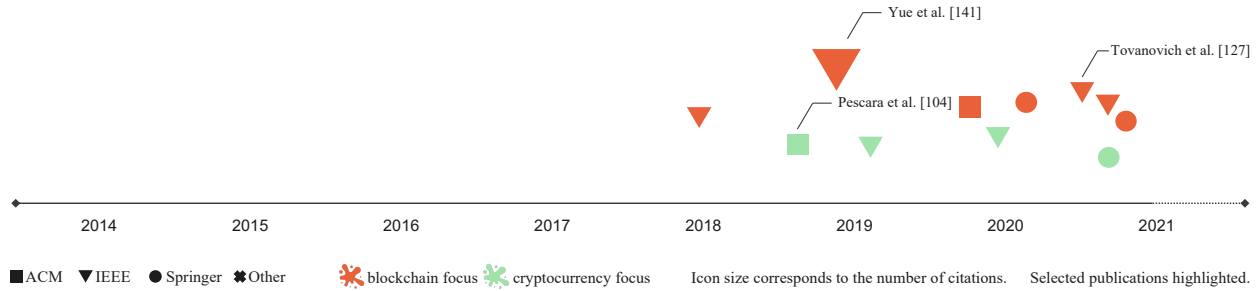

Figure 10: Overview of publications assigned to the *Blockchain: Support Tools* theme.

*4.6.1 Transaction Analytics and Visualization.* Transactions on most blockchain-based networks are public. However, due to the sheer number of transactions and their pseudonymous design it is hard for novices and experts alike to make sense of the data in front of them, which is usually only provided in the form of text [141]. Transaction analytics tools aim to transform this data into a more human-friendly format. Yue et al.'s BitExTract enables its users to gain a better understanding of transactions between large Bitcoin exchanges. Several researchers focus on systems to better visualize connections between Bitcoin addresses. By offering advanced filters and analytics they aim to support law enforcement or make interactions simpler for users [120, 141, 145]. Tovanovich et al. present an extensive review about visualization of blockchain data by surveying existing applications and academic literature [127], which offers an excellent overview of state-of-the-art approaches.

*4.6.2 Development Support Tools.* Another set of publications is dedicated to the improvement of smart contracts development – particularly, to lower the entry bar for developers with less technical expertise through low-code tools. Tan et al. present a prototype for a *visual smart contract construction system* that allows non-programmers to develop smart contracts [125].

Pursuing a similar objective, Weingärtner et al. aim to make smart contract development more accessible for non-computer experts. They present a graphical programming language for the development of legal smart contracts and, in a brief evaluation, collect indicative evidence that people without programming knowledge can use it [135]. Hossain et al. develop a graphical user interface for the Multichain, a cross-chain router protocol, to make it accessible for people from non-technical backgrounds. Their evaluation showed higher efficiency, better user satisfaction, and an increased overall usability when compared to the original command line interface [57].

## 5 DISCUSSION

Our systematic literature review provides an overview of HCI research on blockchain and cryptocurrencies. We aim to synthesize academic work that has evolved around the experiences, socio-technical challenges, and the design knowledge about blockchain applications. In the following, we draw on recent developments within the wider cryptocurrency and blockchain space to discuss overlaps and differences of the progress observed between research and practice.

### 5.1 Recent Developments in the Blockchain Ecosystem

The blockchain ecosystem has experienced fast-paced growth over the last decade [29]. While until recently, Ethereum was the only widely used permissionless blockchain platform supporting decentralized applications, today, several new blockchains for decentralized applications have reached maturity [118]. Ethereum and Bitcoin remain the largest ecosystems, yet newcomers like Solana, Polkadot, and Cosmos boast vibrant developer communities with more than 500 monthly active contributors. Many of these emerging blockchains (e.g. Solana, Polkadot, Terra) even exhibit faster ecosystem growth than Ethereum [118]. What distinguishes many of these new blockchains from Ethereum is a host of different technical innovations aimed at overcoming current limitations, particularly speed, transaction throughput, and expensive fees. Much of the challenge of improving the transaction throughput of a blockchain is related to the so-called *blockchain scalability trilemmma*. In essence, it is assumed that for any particular blockchain its *scalability*, *security*, and *decentralization* are dependent features. An improvement to either one of these properties will negatively affect at least one of the others [95]. While Ethereum, with its sizeable decentralized ecosystem, seems to struggle to deploy the required infrastructural changes to overcome its current limitations, the upcoming challengers act more agile. The ongoing emergence of several blockchain systems in parallel can thus be traced back to an opportune moment to challenge the Ethereum ecosystem and to diverging approaches to balance the scalability trilemma in doing so.

Many believe that this new generation of blockchains, now providing transactions at instantaneous speed and low transaction costs, will herald the third stage of the web. Web 1.0 allowed users on the internet the possibility to *read* content. Web 2.0 introduced the option to *write*, and thus enabled rich interactive internet applications. Powered by blockchain, *web3* now adds the possibility to *own*, create, and distribute digital assets. Many practitioners believe this read-write-own paradigm will enable a new class of internet applications with a significant potential for innovation [10]. First indications of this paradigm shift are the emergence of decentralized finance (DeFi) and non-fungible tokens (NFTs), which by now account for over two-thirds of transactions on the Ethereum blockchain [130] and are a driver for user adoption of Ethereum [24].

Juxtaposing the development of the blockchain and cryptocurrency ecosystem with the available research analyzed in this review reveals several gaps. While many of the issues identified by past HCI research are now being addressed through emerging blockchain platforms and technological improvements, formal validation is outstanding. For example, stablecoins address price volatility, and new application blockchains, enabled by novel consensus algorithms, provide high transaction throughput with low-cost fees. The Ethereum Name Service[12] (ENS) maps alphanumerical wallet addresses to human-readable names, allowing users to easily share their wallets. Emerging gateway services like Infura[13] bridge the gap between blockchains and the web for developers. However, until now, HCI research has overwhelmingly focused on only two large blockchain platforms, Bitcoin and Ethereum. This leaves a gap in understanding the full potential of these new technologies, particularly how we can build interactive, usable, secure, and user-centered blockchain applications.

While some work designed and discussed dedicated mobile applications (e.g. [9, 100, 107, 129]), the majority of decentralized applications (dApps) runs in the web browser. Being the de-facto gateway to web3, browser-based wallets such as Metamask[14] facilitate the interaction with dApps. However, we have not found a single study looking into browser-based wallets, leaving a critical gap in understanding their role for interaction with decentralized applications. This is particularly relevant as the emergence of web3 is accompanied by phenomena challenging human interaction and collaboration on the internet. DeFi, NFTs, and decentralized autonomous organizations (DAOs) are the most widespread examples that have driven recent user adoption. To date, only little research has been conducted around DeFi and NFTs. While research started exploring specific use cases for DAOs from a technical perspective, we have only identified a single paper that examined the specific impact of infrastructural limitations (i.e. fees) on user participation in DAOs. We know little about how people within these decentralized organizations manage the socio-technical challenges arising from the tension between pseudonymity and the need to collaborate and trust each other.

Arguably, it is time for HCI to move beyond Bitcoin, chart into new waters, and explore the increasingly diverse ecosystem of cryptocurrencies and distributed ledger technologies[15] as a whole. The technical advances in the field offer a plethora of opportunities to use blockchain as a design material to experiment with novel forms of interaction design and craft rich and interactive experiences.

---

[12]https://ens.domains/ (last-accessed 2022-2-18)
[13]https://infura.io/ (last-accessed 2022-02-18)
[14]https://metamask.io/ (last-accessed 2022-2-18)
[15]For practitioners and researchers with interest in designing and building with blockchain, we can recommend the following article providing an overview of the unique capabilities of recent blockchain protocols and platforms: https://medium.com/coinmonks/unhyped-comparison-of-blockchain-platforms-679e122947c1 (last-accessed 2022-04-19)

## 5.2 Future Research Agenda

This discussion and its preceding literature review highlight the importance of HCI in the ongoing development of blockchain applications. Over the past 8 years, a diverse research body has been established through the works of many scholars. To conclude this paper, we present five research avenues the HCI and interaction design community may address in the future.

*5.2.1 A better understanding of Blockchain Users.* Existing research shows that blockchain and cryptocurrency users are an increasingly heterogeneous group with different motivations, needs, skills, and experiences. With first works untangling the user base of cryptocurrency existing [2], there remains more work to better understand and segment users. Particularly, the recent emergence of web3, most prominently through DeFi and NFTs, has likely drawn in new users with different motivations and expectations than the early Bitcoin adopters. For example, "Twitter NFT" has emerged as a subculture with its own language (e.g. "gm", "probably nothing", "WAGMI") [108]. Likely the ideology connecting people within this group is quite different from the "True Bitcoiner" ideology reported by Knittel et al. [76, 77] and HCI should continue to aim for a better understanding of the economic context under which people become involved with web3. Contesting borders between the digital and physical world, we have seen examples of virtual groups of people organizing themselves into DAOs to achieve common goals. For example, Constitution DAOs attracted more than 19,000 members in an effort to buy a rare copy of the US constitution [110]. Building on the existing research body about trust, future scholars may explore how these decentralized pseudonymous groups organize themselves, build trust, and maintain it over time.

With diversity and inclusion being longstanding values within the HCI community, another topic to address is the question of why there is such a gender imbalance in the blockchain space. Multiple authors recognize this imbalance in the demographics of their papers, yet none of them attempted to find an explanation. With organizations like Global Women in Blockchain[16] aiming to empower women to engage with the technology, change is happening, and numbers are slowly growing [86]. Being champions of diversity, we urge the HCI community to take an active role in identifying the reasons that hold women back from engaging with the technology and make an effort to change that.

*5.2.2 Generative Interaction Design for Wallets.* Our review shows that existing research has investigated the perception and usability of various cryptocurrency wallets in both qualitative and quantitative studies. Many scholars highlight challenges and propose implications for design – however, these remain largely untested. We identified only three publications [18, 30, 48] implementing wallets or prototyping interfaces. Given that wallets are essential to interact with cryptocurrencies and dApps, future interaction design research is challenged to fill this gap. The ultimate outcome of this strand of research could be a set of validated design heuristics and guidelines specific to cryptocurrencies, as suggested by Voskobojnikov et al. [133]. Against the backdrop of an increasingly diverse blockchain ecosystem, it is likely necessary to explore wallets for different use cases and on different devices to develop these heuristics: Hardware wallets for secure long term storage, exchanges and online wallets for quick access and trading, mobile wallets for payments, and browser-based wallets for interaction with dApps on both desktop and mobile devices.

---

[16]https://globalwomeninblockchain.org/ (last-accessed 2022-2-18)

Assuming a growing integration of blockchain into the web, more and more information will be tied to a specific address. It will be important to design and evaluate educational concepts helping users to update their mental models and overcome misconceptions that otherwise could lead to costly mistakes. To make use of the full benefits promised by blockchain technology, users need to manage their keys on their own. While certainly not desired by all users, exploring ways to safely transition from custodial to self-managed wallets will be important to reduce losses for users who want to. Even though some papers mentioned the positive innovation cryptocurrency has brought to key management (e.g. mnemonics, private keys encoded in 12-word phrases) there was no study in our sample that actively explored this design space. Interaction design can take an active role in developing concepts for key management that nudge users towards secure behavior and provide usable security.

*5.2.3 Moving beyond Bitcoin.* Bitcoin has laid the foundation for cryptocurrency and blockchain adoption, so it is not surprising that the majority of existing research focuses on the use of Bitcoin. However, the cryptocurrency and blockchain space is diversifying with new generations of blockchain platforms, which are being increasingly adopted by users, developers, and the market [118]. This can also be seen in the gradual decline of Bitcoin's dominance [28]. We argue that future research should be confident to move beyond Bitcoin and adopt state-of-the-art blockchains both as a research subject and platform for new designs and innovation. Doing so two directions will be particularly interesting.

First, we suggest to evaluate whether emerging technologies are able to fulfill their promise to overcome the performance and scalability issues identified by literature across the domain. Due to the current focus on technology that was introduced some 6-10 years ago, some of the issues pertaining to cryptocurrencies might be less prevalent or even solved through advancements in the technology today. In particular, the challenges around scalability and fees could be revisited to update the sector's understanding.

The second direction is to explore and prototype with the increasingly specialized set of blockchains as design material: Decentralized application platforms – e.g. layer-1 platforms such as Polkadot, Solana and Cosmos and layer-2 blockchains like Polygon, Avalanche, Terra, or Bitcoin Lightning – offer novel opportunities for interaction design. Development tools for smart contract development have matured over the past years, making it easier to design and build smart contracts and decentralized applications. With their promise for faster transaction speeds at lower costs researchers and designers can chart the design space for truly interactive blockchain applications.

*5.2.4 Engaging with Web3 and Decentralized Applications (dApps).* An increasing number of decentralized applications is being adopted by users [24]. This large variety of new applications offers vast opportunities for HCI to research fundamental socio-technical mechanisms connected to blockchain technology. With new technical and mental models being developed, it is a promising space for service and interaction designers.

Measured by the gas fee burn rates, today around two-thirds of transactions on the Ethereum blockchain can be attributed to either NFTs or DeFi, having superseded the mere monetary transfers [130]. While these application areas have been exhibiting increased adoption by users in recent years, this trend has not been reflected in the amount of research being carried out within HCI. In the case of DeFi, the design of interfaces and support tools could have a substantial influence on user behavior (c.f. [78]). More dynamic, intelligent interfaces could, for example, guide users to make better decisions on complex transactions within decentralized exchanges to avoid transactions being delayed or even intercepted. Elsden et al. [35] envisaged the opportunities of digital ownership on blockchain. With the emergence of NFTs this became a reality. NFTs offer an opportunity to further explore the meaning of digital ownership and could revolutionize how digital content creators design, create, trade, and own digital assets. At the same time, NFTs sparked discussion about the value and uniqueness of digital items that can be easily copied. Nevertheless, more and more people are willing to pay for them and thus derive some benefit from owning them.

With the majority of decentralized applications being consumed through the web browser, there is a need to better understand the role of gateway services. Decentralized applications on web3 frequently do not connect to the blockchain directly but through centralized services like Infura. The role of reintermediation of a disintermediated system raises questions about how to maintain power balances, privacy, and the integrity of data visualized in the actual user interfaces that have so far not been addressed by research.

*5.2.5 Identity on the ledger.* Despite the large public interest, our findings highlighted a significant research gap in HCI around self-sovereign identity management (SSI). SSI has the potential to manage identities in a simple, uncomplicated, trustworthy, and self-reliant way. We would like to encourage the HCI and interaction design community to explore research avenues in this direction. Comparable to an identity document like a passport, web3 opens up opportunities to create virtual identities and reputation that counter-balance the trust challenges [114, 115] in an otherwise pseudonymous system. Aimed at overcoming the need for isolated accounts on every web platform, Sign-In-With-Ethereum[17] allows developers to use the wallet address of a user to authenticate them. While beneficial from the standpoint of privacy and security from a user's standpoint – gone is the need to share e-mail addresses or enter passwords – this arguably raises questions for website operators on how to deal with the loss of information that today is often at the core of internet business models.

Blockchain-based identity extends beyond technical aspects and opens up fundamental questions about how human identity can be expressed in an increasingly digital world. The Ethereum Name Service is the most widely used tool that allows users to connect their wallet address to a human-readable name, comparable to how domain name services (DNS) map names and IP addresses. This seemingly superficial abstraction allows users to establish a shareable and permanent identity to which they can link their online personas. By doing so, they can build a reputation through transactions connected to their addresses that is public to see and easy to verify by others. This phenomenon can already be seen in the context of web3: People are starting to use NFTs as a form of human expression and self-identity on social media. They present themselves through online personas disconnected from their real identities, set NFTs as profile pictures, use them as avatars in video calls (see e.g. huddle01.com[18]), or use the transaction history connected to their wallets as source of reputation (see e.g. POAPs[19]). It remains to be seen in how far self-sovereign identity can prevail against the centralized services that govern the internet today. For HCI, there is an opportunity to chart the designed space of digital identity, connecting the underlying technological constraints with the fundamental human need for the expression of one's identity.

---

[17]https://login.xyz/ (last-accessed 2022-2-18)

# 6 CONCLUSION

This paper presents a systematic literature review of blockchain and cryptocurrency research in HCI. Our analysis includes 99 relevant papers published between 2014 and 2021. We identify six salient themes: 1) the role of trust, (2) understanding motivation, risk, and perception of cryptocurrencies, (3) the usability of cryptocurrency wallets, (4) engaging users with blockchain, (5) using blockchain for application-specific use-cases, and (6) designing support tools for blockchain. We summarize the generated design knowledge, discuss open challenges, and juxtapose the current research body with the changing landscape of emerging blockchain technologies to chart the space for future HCI research. We encourage HCI researcher to better understand blockchain users, take an active approach to designing wallets, adopt new blockchains as design material, engage with web3 and decentralized applications, and explore digital identity. We hope that this paper provides a valuable overview of the current state of blockchain and cryptocurrency research in HCI and that it can act as road map for researchers and practitioners moving forward.

---

[18] https://huddle01.com/ (last-accessed 2022-2-18)
[19] https://poap.xyz/ (last-accessed 2022-2-18)

Table 5: Overview of all publications included in the review.

| Reference | year | Focus | | Blockchain | | | | Contribution Type | | | | | | | | Major Themes | | | | | |
|---|---|---|---|---|---|---|---|---|---|---|---|---|---|---|---|---|---|---|---|---|---|
| | | blockchain | cryptocurrency | bitcoin | ethereum | Other | not specified | empirical (system) | empirical (people) | artifact | method | theory | dataset | literature review | essay | trust | motivation,risk,perc. | wallets | engaging users | specific use cases | support tools |
| [117] | 2014 | - | • | • | - | • | - | - | • | - | - | - | - | - | - | - | • | - | - | - | - |
| [90] | 2015 | • | - | • | - | - | - | • | - | - | - | - | - | - | - | - | - | - | • | - | - |
| [114] | 2015 | - | • | • | - | - | - | - | - | - | • | - | - | - | - | • | - | - | - | - | - |
| [40] | 2016 | - | • | • | - | - | - | • | - | • | - | - | - | - | - | - | - | - | - | • | - |
| [71] | 2016 | - | • | • | - | - | - | - | • | - | - | - | - | - | - | - | • | - | - | - | - |
| [51] | 2016 | - | • | • | - | - | - | - | • | - | - | - | - | - | - | - | • | - | - | - | - |
| [45] | 2017 | • | - | • | - | - | - | - | - | - | - | - | - | • | - | - | - | - | - | • | - |
| [113] | 2017 | • | - | - | • | - | - | • | - | • | - | - | - | - | - | - | - | - | - | • | - |
| [101] | 2017 | • | - | - | • | - | - | - | - | • | - | - | - | - | - | - | - | - | - | • | - |
| [67] | 2017 | - | • | • | - | - | - | • | - | - | - | - | - | - | - | - | - | • | - | - | - |
| [115] | 2017 | - | • | • | - | - | - | - | • | - | • | - | - | - | - | • | • | - | - | - | - |
| [79] | 2017 | - | • | • | - | - | - | - | • | - | - | - | - | - | - | - | • | • | - | - | - |
| [56] | 2017 | - | • | • | - | - | - | - | • | - | - | - | - | - | - | - | - | - | - | • | - |
| [30] | 2017 | - | • | • | - | - | - | - | • | - | - | - | - | - | - | - | • | - | - | - | - |
| [35] | 2018 | • | - | • | • | • | - | - | - | - | - | • | • | - | - | - | - | - | - | • | - |
| [100] | 2018 | • | - | • | - | - | - | • | - | • | - | - | - | - | - | - | - | - | • | • | - |
| [122] | 2018 | • | - | • | - | - | - | • | - | • | - | - | - | - | - | - | - | - | - | • | - |
| [135] | 2018 | • | - | - | • | - | - | • | - | • | - | - | - | - | - | - | - | - | - | - | • |
| [16] | 2018 | • | - | - | • | - | - | • | - | • | - | - | - | - | - | - | - | - | - | • | - |
| [42] | 2018 | • | - | - | - | • | - | • | - | • | - | - | - | - | - | - | - | - | - | • | - |
| [91] | 2018 | • | - | - | - | • | - | • | - | • | - | - | - | - | - | - | - | - | - | • | - |
| [88] | 2018 | • | - | - | - | • | - | • | - | - | - | - | - | - | - | - | - | - | • | - | - |
| [37] | 2018 | • | - | - | - | • | - | • | - | - | - | - | - | - | - | - | - | - | • | • | - |
| [20] | 2018 | • | - | - | - | • | - | - | • | - | - | - | - | - | - | • | - | - | - | • | - |
| [61] | 2018 | • | - | - | - | • | - | - | • | - | - | - | - | - | - | - | - | - | - | • | - |
| [9] | 2018 | • | - | - | - | • | - | - | • | - | - | - | - | - | - | - | - | - | - | • | - |
| [63] | 2018 | - | • | • | • | • | - | - | - | - | - | - | - | - | - | - | • | • | - | — | - |
| [78] | 2018 | - | • | • | • | • | - | - | • | - | - | - | - | - | - | - | • | • | - | - | - |
| [109] | 2018 | - | • | • | - | - | - | • | - | - | - | - | - | - | - | - | - | • | - | - | - |
| [72] | 2019 | • | - | • | - | - | - | • | - | • | - | - | - | - | - | - | - | - | - | • | - |
| [141] | 2019 | • | - | • | - | - | - | - | - | - | - | - | - | - | - | - | - | - | - | - | • |
| [53] | 2019 | • | - | • | - | - | - | - | - | - | • | - | - | - | - | • | - | - | - | - | - |
| [136] | 2019 | • | - | - | • | - | - | • | - | • | - | - | - | - | - | - | - | - | - | • | - |
| [116] | 2019 | • | - | - | • | - | - | • | - | • | - | - | - | - | - | - | • | - | - | • | - |
| [68] | 2019 | • | - | - | • | - | - | • | - | - | - | - | - | - | - | - | - | - | • | - | - |
| [142] | 2019 | • | - | - | - | • | - | • | - | - | - | - | - | - | - | • | - | - | - | • | - |
| [38] | 2019 | • | - | - | - | • | - | - | • | - | - | - | - | - | - | - | - | - | - | • | - |
| [36] | 2019 | • | - | - | - | • | - | - | • | - | - | - | - | - | - | - | - | - | - | • | - |
| [62] | 2019 | • | - | - | - | • | - | - | • | - | - | - | - | - | - | - | - | - | - | • | - |
| [73] | 2019 | • | - | - | - | • | - | - | - | - | - | - | - | - | • | • | - | - | - | - | - |
| [54] | 2019 | • | - | - | - | • | - | - | - | - | - | - | - | - | • | - | - | - | - | • | - |
| [21] | 2019 | • | - | - | - | • | - | - | - | - | - | - | - | - | • | • | - | - | - | - | - |
| [84] | 2019 | • | - | - | - | • | - | - | - | - | - | - | - | - | • | - | - | - | - | • | - |
| [34] | 2019 | • | - | - | - | • | - | - | - | - | - | - | - | - | • | - | - | - | - | • | - |
| [5] | 2019 | - | • | • | - | - | - | • | - | - | - | - | - | - | - | - | • | • | - | - | - |
| [8] | 2019 | - | • | • | - | - | - | • | - | - | - | - | - | - | - | - | • | • | - | - | - |
| [76] | 2019 | - | • | • | - | - | - | - | • | - | - | - | - | - | - | • | • | - | - | - | - |
| [70] | 2019 | - | • | • | - | - | - | - | • | - | - | • | - | - | - | • | • | - | - | - | - |
| [77] | 2019 | - | • | • | - | - | - | - | • | - | - | - | - | - | - | • | • | - | - | - | - |

Table 5 (continued): Overview of all publications included in the review.

| Reference | year | Focus | | Blockchain | | | | Contribution Type | | | | | | | | Major Themes | | | | | |
|---|---|---|---|---|---|---|---|---|---|---|---|---|---|---|---|---|---|---|---|---|---|
| | | blockchain | cryptocurrency | bitcoin | ethereum | Other | not specified | empirical (system) | empirical (people) | artifact | method | theory | dataset | literature review | essay | trust | motivation,risk,perc. | wallets | engaging with users | specific use cases | support tools |
| [26] | 2019 | - | ● | ● | - | - | - | - | ● | - | - | - | - | - | - | ● | - | - | - | - | - |
| [25] | 2019 | - | ● | ● | - | - | - | - | ● | - | - | - | - | - | - | ● | ● | - | - | - | - |
| [120] | 2019 | - | ● | ● | - | - | - | - | - | ● | - | - | - | - | - | - | - | - | - | - | ● |
| [18] | 2019 | - | ● | - | ● | - | - | ● | - | ● | - | - | - | - | - | - | - | - | ● | - | - |
| [60] | 2019 | - | ● | - | - | - | ● | ● | - | ● | - | - | - | - | - | - | - | - | - | ● | - |
| [104] | 2019 | - | ● | - | - | - | ● | ● | - | - | - | - | - | - | - | - | - | - | - | - | ● |
| [64] | 2020 | ● | - | ● | ● | ● | - | ● | - | ● | - | - | - | - | - | - | - | - | - | ● | - |
| [139] | 2020 | ● | - | ● | - | - | - | - | - | ● | - | - | - | - | - | - | - | - | - | - | ● |
| [17] | 2020 | ● | - | - | ● | - | - | ● | - | ● | - | - | - | - | - | - | - | - | - | ● | - |
| [124] | 2020 | ● | - | - | ● | - | - | ● | - | ● | - | - | - | - | - | - | - | - | - | ● | - |
| [129] | 2020 | ● | - | - | ● | - | - | ● | - | ● | - | - | - | - | - | - | - | - | - | ● | - |
| [125] | 2020 | ● | - | - | ● | - | - | ● | - | ● | - | - | - | - | - | - | - | - | - | - | ● |
| [3] | 2020 | ● | - | - | ● | - | - | ● | - | ● | - | - | - | - | - | - | - | - | - | ● | - |
| [19] | 2020 | ● | - | - | ● | - | - | - | - | ● | - | - | - | - | - | - | - | - | - | ● | - |
| [107] | 2020 | ● | - | - | - | - | ● | ● | - | ● | - | - | - | - | - | - | - | - | - | ● | - |
| [143] | 2020 | ● | - | - | - | - | ● | ● | - | - | - | - | - | - | - | ● | - | - | - | - | - |
| [52] | 2020 | ● | - | - | - | - | ● | ● | - | - | - | - | - | - | - | - | - | - | ● | - | - |
| [111] | 2020 | ● | - | - | - | - | ● | ● | - | - | - | - | - | - | - | - | - | - | - | ● | - |
| [103] | 2020 | ● | - | - | - | - | ● | - | ● | - | - | - | - | - | - | - | ● | - | - | - | - |
| [32] | 2020 | ● | - | - | - | - | ● | - | ● | - | - | - | - | - | - | - | - | - | - | ● | - |
| [134] | 2020 | ● | - | - | - | - | ● | - | ● | - | - | - | - | - | - | ● | - | - | - | - | - |
| [128] | 2020 | ● | - | - | - | - | ● | - | - | ● | - | - | - | - | - | - | - | - | - | ● | - |
| [22] | 2020 | ● | - | - | - | - | ● | - | - | - | - | - | - | - | ● | - | - | - | - | ● | - |
| [94] | 2020 | - | ● | ● | ● | ● | - | ● | - | - | - | - | - | - | - | - | ● | - | - | - | - |
| [14] | 2020 | - | ● | ● | ● | ● | - | - | - | ● | - | - | - | - | - | - | - | ● | - | - | - |
| [145] | 2020 | - | ● | ● | - | - | - | ● | - | ● | - | - | - | - | - | - | - | - | - | - | ● |
| [50] | 2020 | - | ● | ● | - | - | - | - | ● | - | ● | - | - | - | - | - | ● | ● | - | - | - |
| [87] | 2020 | - | ● | ● | - | - | - | - | ● | - | - | - | - | - | - | - | - | ● | - | - | - |
| [102] | 2020 | - | ● | ● | - | - | - | - | ● | - | - | - | - | - | - | ● | - | ● | - | - | - |
| [7] | 2020 | - | ● | ● | - | - | - | - | ● | - | - | - | - | - | - | - | - | ● | - | - | - |
| [65] | 2020 | - | ● | - | - | - | ● | ● | - | - | - | - | - | - | - | - | - | ● | - | - | - |
| [6] | 2020 | - | ● | - | - | - | ● | - | ● | - | - | - | - | - | - | - | ● | - | - | - | - |
| [55] | 2021 | ● | - | ● | ● | - | - | ● | - | - | - | - | - | - | - | - | - | - | - | ● | ● | - |
| [127] | 2021 | ● | - | ● | - | - | - | - | - | - | - | - | - | ● | - | - | - | - | - | - | ● |
| [75] | 2021 | ● | - | ● | - | - | - | ● | - | ● | - | - | - | - | - | - | - | - | - | - | ● |
| [123] | 2021 | ● | - | - | ● | - | - | ● | - | - | - | - | - | - | - | - | - | - | - | ● | - |
| [81] | 2021 | ● | - | - | ● | - | - | ● | - | - | - | - | - | - | - | - | - | - | ● | - | - |
| [41] | 2021 | ● | - | - | ● | - | - | ● | - | - | - | - | - | - | - | - | - | ● | - | - | - |
| [1] | 2021 | ● | - | - | ● | - | - | ● | - | - | - | - | - | - | - | - | - | - | - | ● | - |
| [57] | 2021 | ● | - | - | - | ● | - | ● | - | ● | - | - | - | - | - | - | - | - | - | - | ● |
| [11] | 2021 | ● | - | - | - | - | ● | ● | - | ● | - | - | - | - | - | - | - | - | - | ● | - |
| [126] | 2021 | ● | - | - | - | - | ● | - | - | - | - | - | - | ● | - | - | - | - | - | ● | - |
| [74] | 2021 | ● | - | - | - | - | ● | - | - | - | - | - | - | - | ● | - | - | - | ● | - | - |
| [133] | 2021 | - | ● | ● | ● | ● | - | ● | - | - | - | - | - | - | - | - | - | ● | ● | - | - |
| [2] | 2021 | - | ● | ● | ● | ● | - | - | ● | - | - | - | - | - | - | - | - | ● | ● | - | - |
| [48] | 2021 | - | ● | ● | - | - | - | ● | - | ● | ● | - | - | - | - | - | - | - | ● | - | - |
| [49] | 2021 | - | ● | ● | - | - | - | ● | - | - | - | - | - | - | - | - | - | - | ● | - | - |
| [82] | 2021 | - | ● | - | - | - | ● | ● | ● | - | - | - | - | - | - | ● | - | - | - | - | ● |
| [66] | 2021 | - | ● | - | - | - | ● | ● | - | - | - | - | - | - | - | - | - | ● | - | - | - |
| [131] | 2021 | - | ● | - | - | - | ● | - | ● | - | - | - | - | - | - | ● | ● | - | - | - | - |